\documentstyle[preprint,aps]{revtex}
\begin{document}
\draft
\title{Strong and electromagnetic interactions of heavy baryons}
\author{R Delbourgo and Dongsheng Liu}
\address{Physics Department, University of Tasmania,\\
GPO Box 252C Hobart, Australia 7001}
\date{\today}
\maketitle

\begin{abstract}
It is possible to express all the strong and electromagnetic interactions of
ground state hadrons in terms of a single coupling constant and the
constituent quark masses, $m_{ud}\simeq 0.34$ GeV, $m_s \simeq 0.43$ GeV and
$m_c \simeq 1.5$ GeV, by using spin-flavour relativistic supermultiplet
theory. We show that this produces results which are generally accurate to
within 10\%. We thereby predict widths and couplings of recently and
soon-to-be discovered heavy hadrons.
\end{abstract}

\pacs{11.30.Hv, 11.30.Ly, 13.30.-a, 13.40.Hq}

\narrowtext

\section{INTRODUCTION}

%I

It is almost 30 years ago since SU(6) theory~\cite{GR} and its relativistic
generalization~\cite{DSS} was conceived, before even the birth of quantum
chromodynamics (QCD). Nowadays it is largely forgotten that, apart from weak
interactions, it was spectacularly successful at predicting the strong and
electromagnetic decays of hadrons. Further, it was realized in 1966 that the
predictions could only be regarded as `tree-level' or effective interactions
between the hadronic states rather than a fully-fledged description, since
unitarity provided definite corrections which broke the spin-flavour
symmetry. However, thanks to the work of Isgur and Wise~\cite{hqet}, today
the symmetry is envisaged as applying to hadrons at equal velocity
containing one heavy quark, since the QCD Lagrangian possesses such a
symmetry in the heavy mass limit~\cite{eff}. The current description
popularly treats the light meson through chiral perturbation theory even
though previous history indicates that they are equally well described by
spin-flavour symmetry, weak interactions notwithstanding, provided that the
quarks are accorded their constituent masses rather than the current quark
values. In this paper we shall take these constituent or effective masses to
be $m_{ud}\simeq 0.34$ GeV, $m_s \simeq 0.43$ GeV and $m_c \simeq 1.5$ GeV,
values which accord quite well with mass formulae and spin-splittings.

Because a great deal of experimental data has become available since 1966
with which to test relativistic supermultiplet schemes, we shall revisit
some of these early predictions to test how well they pan out and, upon
satisying ourselves that they generally lie within about 10\% of the data,
we will extrapolate to the heavy hadrons where they should be even more
secure according to heavy quark lore. We intend to concentrate on processes
and features that are amenable to experimental testing soon and will avoid
weak decays: an area where understandably most of the recent research on
heavy quarks is focussed, because that is where the bulk of the data is to
be found. The imminent arrival of beauty and charm factories promises an
explosion of results every bit as impressive as the late 60's and early 70's
proved to be for the strange hadronic states, and not purely in the c\={c}
and b\={b} sector.

Instead of relying on tables of Clebsch-Gordan coefficients for the higher
groups, we will base our analysis on a simple multispinor construction which
produces the required symmetry relations from first principles. These states
are tabled in the appendix, listed in terms of the multispinors. It is very
simple to read off the answers as needed or program them into algebraic
computer packages like Maple, to check or actually determine the requisite
matrix elements. This procedure now goes under the name of the `trace
formula'~\cite{TF}.

In the next section we shall set out the formalism. Our treatment of the
quarks is deliberately naive as we wish to see how much one can learn simply
by boosting up from rest the composite wave-functions describing the
hadrons, without taking account of any additional, finer effects. Our
comparisons with the experimental data are given in the following three
sections and the results indicate that subtler QCD corrections are rather
minor, which is puzzling given our present knowledge of QCD.

\section{Multispinor states}

We make the assumption, common to all quark models, that the hadrons are
bound colourless S-wave states, of quark and antiquark for mesons, of three
quarks for baryons. We take it that these hadrons consist of the various
quarks moving in tandem, {\em with the same velocity} and, in keeping with
our naive perspective, we shall neglect virtual gluons by supposing that
their main function, apart from keeping the pieces together, is to give the
quarks their composite (dynamical) masses. Neglecting the relative motion
between quarks, which must of course average to zero, the states can be
expressed as products of multispinors. We therefore represent the rest frame
baryonic states by $\Psi_{(ABC)}$, with $2N(N+1)(2N+1)/3$ components, where $%
N$ is the number of flavours and $A\equiv\alpha a$. $a$ stands for the
flavour index and $\alpha$ is the spinor index; $\alpha$ has only 2
effective components because of the on-shell spinor equation, which reads $%
(\gamma\cdot v-1)u(v)=0$.

We can decompose the multispinor into SU($N$)$\times$SU(2) components in the
traditional way:
\begin{equation}
\Psi_{(ABC)} = \psi_{(abc)(\alpha\beta\gamma)} + \frac{\sqrt{2}}{3}
\left(\psi_{[ab]c[\alpha\beta]\gamma} + \psi_{[bc]a[\beta\gamma]\alpha} +
\psi_{[ca]b[\gamma\alpha]\beta} \right).
\end{equation}
Our normalization is fixed by
\[
\bar{\Psi}^{(ABC)}\Psi_{(ABC)}=\bar{\psi}^{(abc)(\alpha\beta\gamma)}
\psi_{(abc)(\alpha\beta\gamma)}+ \bar{\psi}^{[ab]c[\alpha\beta]\gamma}
\psi_{[ab]c[\alpha\beta]\gamma},
\]
and one may verify that the total number of components match up: there are
the spin 3/2 SU(2) spinors, symmetric in flavour indices $(abc)$, having $%
N(N+1)(N+2)/6$ components, as well as the spin 1/2 SU(2) spinors of mixed
symmetry $[ab]c$, with $N(N^2-1)/3$ components. See the appendix for extra
details, listing the multispinors relations to the particle states
themselves. A similar treatment, when applied to the mesons, yields the
vector-pseudoscalar supermultiplet:
\[
\Phi_A^B = \delta_a^b\phi_{5\alpha}^\beta + \vec{\sigma}_a^b\cdot\vec{\phi}%
_\alpha^\beta.
\]

Then, upon boosting up the quarks from rest, the wavefunctions assume their
relativistic form ($v$ denotes the incoming hadron 4-velocity):
\begin{eqnarray}
\Psi_{(ABC)}(v)&=&[P_{+v}\gamma_\mu C]_{\alpha\beta} u^\mu_{\gamma(abc)}(v) +
\nonumber \\
& & \frac{\sqrt{2}}{3} \left([ P_{+v}\gamma_5C]_{\alpha\beta}u_{[ab]c%
\gamma}(v)+ [ P_{+v}\gamma_5C]_{\beta\gamma}u_{[bc]a\alpha}(v)+ [
P_{+v}\gamma_5C]_{\gamma\alpha}u_{[ca]b\beta}(v)\right).
\end{eqnarray}
\begin{equation}
\Phi_A^B(q) = \left[ \mu P_{+v}(\gamma_5\phi_{5a}^b(q)- \gamma^\nu\phi_{\nu
a}^b(q)) \right]; \qquad q=\mu v,
\end{equation}
where $P_{+v} \equiv (1+\!\not{\!}\!v)/2$ is the positive energy projector.
Of course the vector fields $u_\mu$ and $\phi_\mu$ obey the constraints, $%
\gamma^\mu u_\mu = v^\mu u_\mu = v^\mu\phi_\mu = 0$.

This much is a direct generalization from SU(6) to SU(2$N$) of the old
treatment. Now historically the quarks were given the same mass---this was
one of the criticisms of the early work---but that assumption is quite
unnecessary as we have learned from heavy quark theory. All one needs to
appreciate is that the quarks have to be travelling with the same velocity,
so that the formulae (2) and (3) apply perfectly well to unequal mass quarks~%
\cite{HKT}. Therefore one can readily substitute $p/m$ for $v$, where $p$ is
the total 4-momentum of the hadron and $m$ is its total mass, without going
wrong.

The processes which we shall examine, including the charmed and bottom
hadrons, have their origin in the strong three-point vertices
\begin{equation}
{\cal L} = F\Phi(q_1)\Phi(q_2)\Phi(q_3) + G\Psi(p^{\prime})\Phi(q)\Psi(p),
\end{equation}
where $F$ and $G$ are `universal' coupling constants. With our convention, $%
\Phi$ has mass dimensions [M]$^2$ and $\Psi$ has dimension [M]$^{3/2}$,
because the component fields possess the conventional dimensions of Fermi
and Bose fields. Therefore $G\sim$ [M]$^{-1}$ and $F \sim$ [M]$^{-2}$ are
dimensionful couplings and we will be faced with interpreting them before
comparing our results with $physical$ amplitudes and decay rates. The point
is that the naive view which we are adopting takes the hadron mass as the
sum of the constituent masses (spin-splitting being neglected in the first
instance); this is sometimes a far cry from the physical mass and we cannot
gloss over this problem.

The electromagnetic interactions in Section V will be handled through the
vector dominance model---albeit with some finesse---and thus follow from the
strong vertices above. Whether we are dealing with pseudoscalar or vector
mesons, the subsidiary conditions ensure that there is an overall factor of
the sum of the participating hadron masses multiplying the couplings $F$ and
$G$. Consequently we shall regard dimensionless $g=3G\Sigma/4$, where $\Sigma
$ is the sum of the masses as the proper universal meson-baryon coupling and
$f=F\mu\Sigma$ as the proper universal meson-meson coupling, from the point
of view of the rest frame SU(2$N$)$\times$SU(2$N$) symmetry. The
consequences of this are explained shortly.

\section{Relating the strong interactions}

To uncover the relations between the strong interactions of the spin
components, one only needs to insert the expansions (2) and (3) into (4) and
take traces as required by the spinor algebra. This mechanical process leads
to the following effective interactions:

\begin{description}
\item  {\underline{$1^- \rightarrow 0^- + 0^-$}}
\begin{equation}
{\cal L}_{311} = \frac{1}{2}f\left( (q_2-q_3)^\lambda
[\phi_\lambda(q_1)\phi_5(q_2)\phi_5(q_3)]_- + {\rm 2~cyclic~perms~in~}q
\right),
\end{equation}
where $[XYZ]_- \equiv X_a^b[Y_b^c Z_c^a - Z_b^c Y_c^a]$ is the antisymmetric
flavour combination, consistent with Bose statistics.

\item  {\underline{$0^- \rightarrow 1^- + 1^-$}}
\begin{equation}
{\cal L}_{133} = f\left( \epsilon_{\mu\nu\rho\sigma} q_2^\rho q_3^\sigma[%
\phi_5(q_1)\phi^\mu(q_2)\phi^\nu(q_3)]_+/\mu + {\rm 2~cyclic~perms~in~} q
\right),
\end{equation}
where $[XYZ]_+ \equiv X_a^b[Y_b^c Z_c^a + Z_b^c Y_c^a]$ is the symmetric
flavour combination; this also is in keeping with Bose symmetry.

\item  {\underline{$1^- \rightarrow 1^- + 1^-$}}
\begin{eqnarray}
{\cal L}_{333}&=&\frac{1}{2}f\left([(q_2-q_3)_\lambda\eta_{\mu\nu}
+(q_3-q_1)_\mu\eta_{\nu\lambda} +(q_1-q_2)_\nu\eta_{\lambda\mu} + \right.
\nonumber \\
& &\left. (q_2-q_3)_\lambda(q_3-q_1)_\mu(q_1-q_2)_\nu/6\mu^2]
[\phi^\lambda(q_1)\phi^\mu(q_2)\phi^\nu(q_3)]_+ + q{\rm ~perms}\right),
\end{eqnarray}
where we have taken the vectors to possess common mass $\mu$. Notice the
similarity of the first part of this expression to the Yang-Mills vertex.

\item  {\underline{$1/2^+ \rightarrow 1/2^+ + 0^-$}}
\begin{equation}
{\cal L}_{221} = \frac{1}{2}g(1+v\cdot v^{\prime}) [\bar{u}%
(p^{\prime})\gamma_5\phi_5(q)u(p)]_{D-S+2F/3},
\end{equation}
where the $F,D,S$ combinations correspond the internal symmetry
combinations:
\begin{equation}
F + 3S \equiv [3\bar{u}^{[bc]a}\phi_a^d u_{[bc]d} + \bar{U}%
^{(bc)a}\phi_a^dU_{(bc)d}]/4,
\end{equation}
\begin{equation}
D - 3S \equiv [\bar{u}^{[bc]a}\phi_a^d u_{[bc]d} - \bar{U}%
^{(bc)a}\phi_a^dU_{(bc)d}]/4,
\end{equation}
\begin{equation}
{\rm and~}\qquad U_{(bc)a} \equiv u_{[ab]c} + u_{[ac]b},
\end{equation}
hailing from SU(3) days. The multispinor $U$ possesses mixed symmetry too;
instead of being antisymmetric in its first two indices like $u$, it is
symmetric in them. Just like $u$, $U$ obeys the cyclicity relation
\[
U_{(ab)c} + U_{(bc)a} + U_{(ca)b} = 0.
\]

\item  {\underline{$1/2^+ \rightarrow 1/2^+ + 1^-$}}

Here we express the interactions in terms of the electric and magnetic form
factor combinations, which multiply the vectors $E_\lambda \equiv
(v+v^{\prime})_\lambda/2$ and $M_\lambda \equiv
\epsilon_{\lambda\kappa\mu\nu}\gamma^\kappa\gamma_5v^\mu v^{\prime\nu}/2$
respectively:
\begin{equation}
{\cal L}_{223} = g\left( \frac{\mu}{2m}[\bar{u}(p^{\prime})E_\lambda
\phi^\lambda u(p)]_{F+3S} + [\bar{u}(p^{\prime})M_\lambda\phi^\lambda
u(p)]_{D-S+2F/3} \right).
\end{equation}
The significant point is that the two form factors (electric and magnetic,
directly associated with helicity amplitudes) are related and the overall
coupling is connected to the pseudoscalar interaction.

\item  {\underline{$3/2^+ \rightarrow 1/2^+ + 0^-$}}

There is but one possible internal index contraction and one gets the
interaction,
\begin{equation}
{\cal L}_{421}=g\bar{u}^{[ab]c}(p^{\prime})v^{\prime\nu}\phi_{5a}^d
u_{(bcd)\nu}(p)/\sqrt{2}
\end{equation}
where the incoming spin 3/2 particle is a Rarita-Schwinger spinor carrying
momentum $p$, {\em symmetric} in its internal indices.

\item  {\underline{$3/2^+ \rightarrow 1/2^+ + 1^-$}}

In general there would be three independent transition amplitudes here but
the spin-flavour symmetry relates them all via the effective coupling,
\begin{equation}
{\cal L}_{423}=g\epsilon_{\kappa\lambda\mu\nu}v^\mu v^{\prime\nu}\bar{u}
^{[ab]c}(p^{\prime})\phi^{\kappa d}_a u^\lambda_{(bcd)}/\sqrt{2}.
\end{equation}
The significance of this will become apparent when we study the radiative
decays of the excited baryons.

\item  {\underline{$3/2^+ \rightarrow 3/2^+ + 0^-$}}

In this case we would normally expect two independent couplings but they
become united in
\begin{equation}
{\cal L}_{441}=\frac{3}{4}g\left(\eta_{\mu\nu}(1+v\cdot v^{\prime})- v_\mu
v^{\prime}_\nu\right) \bar{u}(p^{\prime})^{(abc)\mu}\gamma_5\phi_{5a}^d
u_{(bcd)}^\nu(p).
\end{equation}
It is much harder to obtain data that tests this relation between the
couplings. However the internal index contraction is at least unique.

\item  {\underline{$3/2^+ \rightarrow 3/2^+ + 1^-$}}

In this case we should expect five independent form factors but they all
collapse into
\begin{equation}
{\cal L}_{443}=\frac{3}{2}g\left( \eta_{\mu\nu}-v_\mu v^{\prime}_\nu
/(1+v\cdot v^{\prime})\right) \bar{u}^{(abc)}(p^{\prime})[(\mu/2m)E_\lambda
+ M_\lambda]u_{(bcd)}(p)\phi^{d\lambda}_a.
\end{equation}
Fortunately there is some experimental data with which to check this
interaction.
\end{description}

\section{Testing the strong interactions}

Because our interactions (5) - (16) apply purely to strong interactions, the
data for checking them out is somewhat limited. We need to look at processes
where the couplings are readily extracted either directly from strong decays
or else from residues of dominant poles in scattering processes. If we
concentrate first on the strong decays, there is considerable data on the
widths of the vector mesons and on the strange baryonic excitations. However
there is little information about the charmed mesons and baryons and what
exists is rather sensitive to the masses of the charmed and bottom excited
states~\cite{KPK}. In some instances the masses are not yet well-determined
so we shall provide a range of predictions, depending on what we assume for
the masses, with a little nous from mass formulae.

The results concerning purely mesonic processes have been published
elsewhere~\cite{JD} so we shall only summarise the findings here. We make
the simplifying approximation that
\[
\phi \simeq s\bar{s}, \qquad\omega\simeq (u\bar{u}+d\bar{d})/\sqrt{2},
\qquad \psi\simeq c\bar{c}
\]
for $1^-$ mesons, but pay proper heed to the mixing angles for $0^-$ states.
Vector meson decays into two pseudoscalars indicate that the corresponding
coupling constant $g_{VPP} = f$ varies slowly with the mass. This is not
altogether surprising from the point of view of heavy quark symmetry, since $%
f$ multiplies a momentum factor, according to (5). Rewriting in terms of
velocities, we anticipate some mass dependence, via a quark loop for
instance; since this is typically governed by the sum of the masses as we
have seen, it suggests we should divide out the mass factor and look for the
constancy of the ratio $g_{VPP}/\sum\mu$ in those processes. The data seems
to bear out this guess fairly well: for $\rho\pi\pi$, K$^*$K$\pi$, $\phi$%
K\=K decays, $g_{VPP}$ equals $4.25, 4.57$ and $4.90$, respectively.
Correspondingly, the mass sum ratios $3m_{ud}, 2m_{ud}+m_s, m_{ud}+2m_s$
provide the ratios 1.02, 1.11 and 1.20 (using the constituent quark masses
mentioned in the introduction) and seem to account for the SU(3) variation
of $g_{VPP}$. Extrapolating to the charmed decays D$^*$D$\pi$, we would
expect $g_{VPP}$ here to equal something like $4.25%
\times(m_c+2m_{ud})/3m_{ud} \simeq 8.9$, which lies below the experimental
bound of $10.2$ but will surely be tested before very long.

Electromagnetic decays offer more clues if one is prepared to apply vector
dominance concepts; we shall discuss those processes presently. Meanwhile,
turning to strong baryon decays, there is a wealth of information from the
spin 3/2 sector. Aside from Clebsch-Gordan coefficients, which can be read
off from the tables at the end, an interaction like (13) leads to a decay
width,
\begin{equation}
\Gamma = \Delta^3g^2(1+v\cdot v^{\prime})/96\pi m^4m^{\prime},
\end{equation}
where $\Delta(m^{\prime},m,\mu) \equiv \sqrt{[m^2-(m^{\prime}+%
\mu)^2][m^2-(m^{\prime}-\mu)^2]}$ is the standard triangle function,
proportional to the magnitude of the decay product three-momentum in the
rest-frame of the decaying particle (mass $m$). After extracting the
physical phase space factors from (17) we may determine the coupling $g$ for
a variety of decays. The results are amazingly constant: all of the decays $%
\Delta \rightarrow {\rm N}\pi, \Sigma^*\rightarrow\Lambda \pi,
\Sigma^*\rightarrow \Sigma\pi$ and $\Xi^* \rightarrow\Xi\pi$, yielding $%
g\simeq 21$, to within 1\%! This encourages us to predict the widths for the
charmed counterparts, $\Sigma_c^*$ and $\Xi_c^{*0}$, provided the
participating masses are precisely known, which they are not. As $%
m(\Sigma_c^*)$ varies from 2.50 GeV to 2.54 GeV the width $\Gamma(\Sigma_c^*
\rightarrow \Lambda_c\pi) \sim 4.5$ to 8.5 MeV, is what we would predict;
the favoured mass and width are 2.53 GeV and 7.1 MeV. Similarly, as $%
m(\Xi_c^{*0})$ runs from 2.62 to 2.65 GeV, we predict that $%
\Gamma(\Xi_c^{*0} \rightarrow \Xi_c\pi)$ will vary between 0.10 MeV and 0.85
MeV, the most likely value being about 0.68 MeV, corresponding to $%
m(\Xi_c^{*0}) = 2.645$ GeV.

One other strong charmed decay is that of the spin 1/2 particle, $\Sigma_c
\rightarrow\Lambda_c\pi$, but before we consider that, let us examine some
better known couplings that follow from pole dominance or dispersion
relations in strong scattering processes. First and foremost there is the
on-shell pion nucleon coupling ($g_{\pi^0pp}$) which is predicted to equal
\[
g_{\pi NN} = g(1-m^2_\pi/4m_N^2)\times 5/6\sqrt{2} \simeq 12.4,
\]
which can be compared with the known value 13.4: a 10\% error seems quite
reasonable considering the extrapolation involved here. Similarly the kaon
couplings are predicted to be
\[
g_{KN\Sigma}=g\frac{(m_N+m_\Sigma)^2-m_K^2}{4m_Nm_\Sigma}  \times \frac{1}{6%
\sqrt{2}} \simeq 2.4 ,
\]
\[
g_{KN\Lambda}=g\frac{(m_N+m_\Lambda)^2-m_K^2}{4m_Nm_\Lambda}  \times \frac{%
\sqrt{6}}{4} \simeq 12.2 .
\]
The information from KN scattering (which is very sensitive to how the
dispersion integrals are evaluated) concentrates on the quantity $%
(g_{KN\Lambda}^2 + 0.85 g_{KN\Sigma}^2)/4\pi$ and gives the range 9 - 17 for
its value. Our prediction of 12.3 lies comfortably within that range.

Moving up to the $\Sigma_c$, the model predicts
\[
g_{\pi{\Lambda_c}{\Sigma_c}}=g\frac{(m_{\Lambda_c}+m_{\Sigma_c})^2 -m_\pi^2}{%
4m_{\Lambda_c}m_{\Sigma_c}\sqrt{6}} \simeq 8.6
\]
and in turn leads to a strong decay width prediction,
\[
\Gamma(\Sigma_c \rightarrow \pi\Lambda_c) \simeq 28 {\rm ~keV} .
\]
Unfortunately the present data tables do not quote a reliable value for
that. The situation is much worse for the bottom mesons and it will probably
be a good while before any sensible numbers are forthcoming for those states.

Before leaving strong interactions, it is worth making some brief remarks
about the vector meson couplings to the baryons. These are obtained from
(12) and include the $\rho$ meson charge coupling. At zero momentum
transfer, $g_\rho$ is related to the pion coupling through
\[
\frac{g_{\pi NN}}{g_{\rho NN}} = \frac{5}{3}\left(\frac{2m}{2\mu}\right)
\left(1-\frac{m_\pi^2}{4m_N^2}\right)  \simeq 5,
\]
upon substituting $m=3m_{ud}$ and $\mu=2m_{ud}$. This gives approximately $%
g_{\rho NN} \simeq g_{\pi NN}/5 \simeq 2.7$, agreeing fairly well with
isospin universality of $\rho$ couplings, which requires that $g_{\rho pp} =
g_{\rho \pi\pi}/2 \simeq 3$. Although we have little direct evidence for
other strong vector couplings to other baryonic states, we do have a large
pool of data on electromagnetic interactions. So we turn to this next.

\section{Relating and testing the electromagnetic interactions}

As mentioned in the introducton, we shall use the vector dominance model
when coupling the photon to the hadrons. In principle we must couple the
photon to all possible $1^{--}$ vector mesons, and this could include the $%
\ell=2$ excitations of the ground state mesons, not to mention radial
excitations. However as these have considerably higher mass than the ground
state particles, it is sufficient for our purpose to mediate the
electromagnetic interaction by the $\ell=0$ states, namely the meson
supermultiplet itself. We believe that it will not greatly damage the
accuracy of our evaluations which are relatively crude anyhow. Now, the
normal procedure is to take the matrix element of the electromagnetic
current $J$ to be
\[
\langle V(k)|J_\lambda^{{\rm em}}|0 \rangle =
e\epsilon_\lambda^*(k)\mu_V^2/g_V,
\]
where $g_V$ is the strong coupling of the vector meson V to the hadrons. Of
course, because we are assuming flavour symmetry, we have $3g_\rho =
g_\omega = -g_\phi = 2g_\psi$ for any hadron.

The strong current is a matrix in flavour space $J_a^d$ and we need only
select the charge projection, $(2J_1^1-J_2^2-J_3^3+2J_4^4)/3$ to ascertain
the relevant part of the strong interaction. However there is one subtle
point about our application of the vector dominance model (VMD) which is
worth pointing out. It has to do with the question of which form factors are
dominated by the vector meson pole, because that choice can make a
substantial difference to the results.

Suppose for instance that we write the strong vector current element in the
traditional manner,
\[
X_\lambda = g\bar{u}(p^{\prime})[\gamma_\mu F_1 +
i\sigma_{\lambda\kappa}q^\kappa F_2]u(p).
\]
Then if were to apply VMD blindly, the electromagnetic current would be $e
X_\lambda/(1 - q^2/\mu^2)$, where $F_1, F_2$ are evaluated on the meson mass
shell ($q^2 = \mu^2$). However if one expresses the strong vector current
element in the alternative way,
\[
Y_\lambda = g\bar{u}(p^{\prime})[E_\lambda F_E + M_\lambda F_M]u(p),
\]
then one may contemplate another VMD version for the electromagnetic current
at non-vanishing momentum transfer, viz. $Y_\lambda/(1-q^2/\mu^2),$ where $%
F_E, F_M$ are worked out on the meson shell. To appreciate the difference,
consider the identity,
\[
i\bar{u}^{\prime}\sigma_{\lambda\kappa}q^\kappa u = \bar{u}^{\prime}[q^2
E_\lambda + 4m^2 M_\lambda]u\times\frac{2m}{4m^2-q^2}.
\]
There is substantial difference between applying VMD to the left-hand-side
(ie multiplying by $\mu^2/(\mu^2-q^2)$) and doing the same {\em at the meson
pole} on the right-hand-side. Therefore we must declare how we propose to
handle this. Because the Sachs form factors $F_E,F_M$ are directly related
to helicity amplitudes and are physically proportional to one another, we
will apply VMD to the electric-magnetic decomposition. This choice then
dictates that the isovector electromagnetic interaction between equal mass
fermions, say, is
\begin{equation}
\langle v^{\prime}|J_\lambda|v\rangle = \frac{1}{2} e\bar{u}%
^{\prime}[E_\lambda + (2m/\mu_V) M_\lambda]u \times \frac{1}{1-q^2/\mu_V^2}.
\end{equation}
Similarly for the isoscalar contribution. The method predicts that the
magnetic moment is $2m/\mu$ in magnetons corresponding to that particle,
times a characteristic Clebsch-Gordan coefficient. Since it is measured in
quark magnetons $e/m$, we can say that the magnetic moment is given as $e/\mu
$ magnetons, where $\mu$ will vary with the mediating meson mass (namely the
sum of its quark constituents). One of the immediate consequences is that
the proton magnetic moment, in nucleon magnetons, equals $m_{{\rm proton}%
}/m_{ud} \simeq 2.75$. More generally we may calculate the magnetic moment
of the spin 1/2 baryons through the linear combination $D-S+2F/3$ arising in
the sum of the components $(2J_1^1-J_2^2-\frac{m_{ud}}{m_s}J_3^3+2\frac{%
m_{ud}}{m_c}J_4^4)/3,$ multiplied by the proton magnetic moment. We have
collected these results in Table IV in the Appendix and also listed the
experimental values for comparison. All in all, the fit is reasonable,
bearing in mind that calculating magnetic moments is a delicate business and
that we have no parameters apart from constituent quark masses, which are
already fixed! The worst prediction is for $\Xi^0$ which is out by 20\%. The
future will produce determinations of moments for charmed and maybe even
bottom baryons, but for the present we must remain ignorant about the
validity of the our predictions for them.

Of course we also have predictions for the spin 3/2 baryons and for
electromagnetic transition elements (3/2 to 1/2), but the data are limited.
Of the excited baryons the only estimated magnetic moment is for the $\Delta
$ resonance. The Particle Data Group~\cite{PD} state that the $\Delta ^{++}$
moment lies between about 4 and 7, while we (really SU(4)) predict that it
equals 5.5; not a very stringent test. However a lot more is known about the
electromagnetic $\Delta ^{+}$-p transition: here one finds the decay rates
expressed in terms of 3/2 and 1/2 helicity amplitudes. The absolute
magnitude of the width $\Gamma _{\Delta ^{+}p\gamma }=0.78$ MeV, implies $%
g_{\Delta p\gamma }\simeq 0.69$ while the supermultiplet prediction is $%
\sqrt{6}e\simeq .73\pm .04$, which is satisfactory. Furthermore, from (16)
one may work out the ratio between the two helicity amplitudes to be $%
S^{3/2}/S^{1/2}=\sqrt{3}:1$. The experimental ratio being $1.82\pm .10$,
this is another good prediction. Unfortunately there is a dearth of data for
transition elements between the strange baryons, except for the transition
moment $\Sigma -\Lambda $ which is quoted in the Table IV. But the situation
is sure to change with time.

\section{Conclusions}

We have seen that all the main features of strong and elctromagnetic
interactions can be understood by relativistically boosting up from rest
spin-flavour symmetric vertices. Apart from the very odd case, all the
results can be described by just {\em one} coupling constant $g$ and three
effective constituent masses for the quarks. They are generally correct to
within 10\%, and often they are better than that. This puts the lie to the
claim that the light meson sector should be handled differently from the
heavy quark sector, although we would be the first to admit that it is not
easy to understand why. After all, the nonstrange and quark dynamical masses
$\sim 300$ to 450 MeV are comparable to the QCD mass scale $\Lambda$.

We have stayed away from weak interactions, because it is necessary to
comprehend how the weak bosons Z and W link with the strong supermultiplets.
While one can see how the vector components of the weak current can be
dominated by the $\ell=0$ mesons, the axial component should couple to the
excited $\ell=1$ meson supermultiplet; this brings in a new, independent
coupling constant. (A proper quark model will relate this to the ground
state coupling of course.) Thus $g_V$ and $g_A$ are {\em distinct} couplings
according to our perspective and their ratio is not given by 5/3 via the
axial-pseudoscalar D/F ratio, as is commonly stated. The bulk of the recent
research activity has naturally been focussed on weak decays, because these
channels predominate, not strong nor electromagnetic channels. We therefore
intend to generalise the work presented in this paper to those processes, as
the next logical step and see how far we can go with only one extra strong
vertex associated with the first orbital excitation of the meson
supermultiplet.

\acknowledgments
We would like to express our thanks to the Australian Research Council who
have supported this research through a grant. We also are indebted to Dr.
Thompson for helpful feedback on our manuscript.

\begin{table}[tbp]
\caption{Mixed symmetry ($\Lambda-$type) states $u_{[ab]c}$ associated with
the spin 1/2$^+$ baryons. Multispinors are antisymmetric in $[ab]$ from
which fact other states are immediately deduced.}
\label{Table 1}
\begin{tabular}{l|cccc}
$ab \downarrow \quad c \rightarrow$ & 1 & 2 & 3 & 4 \\
\tableline 12 & $p/\sqrt{2}$ & $n/\sqrt{2}$ & $\Lambda/\sqrt{3}$ & $%
\Lambda_c^+/\sqrt{3}$ \\
13 & $\Sigma^+/\sqrt{2}$ & $\Sigma^0/2 + \Lambda^0/2\sqrt{3}$ & $-\Xi^0/%
\sqrt{2}$ & $-\Xi_c^+/\sqrt{3}$ \\
14 & $\Sigma_c^{++}/\sqrt{2}$ & $\Sigma_c^+/2 + \Lambda_c^+/2\sqrt{3}$ & $%
\Xi_c^{+\prime}/2-\Xi_c^+/2\sqrt{3}$ & $-\Xi_{cc}^{++}/\sqrt{2}$ \\
23 & $\Sigma^0/2 - \Lambda/2\sqrt{3}$ & $\Sigma^-/\sqrt{2}$ & $-\Xi^-/\sqrt{2%
}$ & $-\Xi_c^0/\sqrt{3}$ \\
24 & $\Sigma_c^+/2 - \Lambda_c^+/2\sqrt{3}$ & $\Sigma_c^0/\sqrt{2}$ & $%
\Xi_c^{\prime 0}/2-\Xi_c^0/2\sqrt{3}$ & $-\Xi_{cc}^+/\sqrt{2}$ \\
34 & $\Xi_c^{\prime +}/2+\Xi_c^+/2\sqrt{3}$ & $\Xi_c^{\prime 0}/2 + \Xi_c^0/2%
\sqrt{3}$ & $\Omega_c^{\prime 0}/\sqrt{2}$ & $\Omega_{cc}^{++}/\sqrt{2}$%
\end{tabular}
\end{table}

\begin{table}[tbp]
\caption{Alternative mixed symmetry ($\Sigma-$type) states $U_{(ab)c}$
associated with the spin 1/2$^+$ baryons. Multispinors are now symmetric in $%
(ab)$ whereupon other states are immediately deduced. Multispinors with
equicomponent indices $U_{(aa)c} = 2u_{[ca]a}$ can be read off from Table I.}
\label{Table 2}
\begin{tabular}{l|cccc}
$ab \downarrow \quad c \rightarrow$ & 1 & 2 & 3 & 4 \\
\tableline 12 & $p/\sqrt{2}$ & $-n/\sqrt{2}$ & $-\Sigma^0$ & $-\Sigma_c$ \\
13 & $\Sigma^+/\sqrt{2}$ & $\Sigma^0/2 -\sqrt{3}\Lambda^0/2$ & $\Xi^0/\sqrt{2%
}$ & $\Xi_c^{\prime +}$ \\
14 & $\Sigma_c^{++}/\sqrt{2}$ & $\Sigma_c^+/2+\sqrt{3}\Lambda_c^+/2$ & $%
\Xi_c^{+\prime}/2+\sqrt{3}\Xi_c^+/2$ & $\Xi_{cc}^{++}/\sqrt{2}$ \\
23 & $\Sigma^0/2+\sqrt{3}\Lambda/2$ & $\Sigma^-/\sqrt{2}$ & $-\Xi^-/\sqrt{2}$
& $-\Xi_c^{\prime 0}$ \\
24 & $\Sigma_c^+/2 +\sqrt{3}\Lambda_c^+/2$ & $\Sigma_c^0/\sqrt{2}$ & $%
\Xi_c^{\prime 0}/2+\sqrt{3}\Xi_c^0/2$ & $\Xi_{cc}^+/\sqrt{2}$ \\
34 & $\Xi_c^{\prime +}/2-\sqrt{3}\Xi_c^+/2$ & $\Xi_c^{\prime 0}/2 - \sqrt{3}%
\Xi_c^0/2$ & $\Omega_c^{\prime 0}/\sqrt{2}$ & $-\Omega_{cc}^{++}/\sqrt{2}$%
\end{tabular}
\end{table}

\begin{table}[tbp]
\caption{Symmetric states $u_{(abc)}$ associated with the spin 3/2$^+$
baryons. Asterisked states in the table are obviously obtainable from the
other entries via the complete symmetry in flavour indices.}
\label{Table 3}
\begin{tabular}{l|cccc}
$ab \downarrow \quad c \rightarrow$ & 1 & 2 & 3 & 4 \\
\tableline 11 & $\Delta^{++}$ & $\Delta^+/\sqrt{3}$ & $\Sigma^{*+}/\sqrt{3}$
& $\Sigma_c^{*++}/\sqrt{3}$ \\
12 & * & $\Delta^0/\sqrt{3}$ & $\Sigma^{*0}/\sqrt{6}$ & $\Sigma_c^{*+}/\sqrt{%
6}$ \\
13 & * & * & $\Xi^{*0}/\sqrt{3}$ & $-\Xi_c^{*+}/\sqrt{6}$ \\
14 & * & * & * & $\Xi_{cc}^{*++}/\sqrt{3}$ \\
22 & * & $\Delta^-$ & $\Sigma^{*-}/\sqrt{3}$ & $\Sigma_c^{*0}/\sqrt{3}$ \\
23 & * & * & $\Xi^{*-}/\sqrt{3}$ & $\Xi_c^{*0}/\sqrt{6}$ \\
24 & * & * & * & $\Xi_{cc}^{*+}/\sqrt{3}$ \\
33 & * & * & $\Omega^-$ & $\Omega_c^{*0}/\sqrt{3}$ \\
34 & * & * & * & $\Omega_{cc}^{*+}/\sqrt{3}$ \\
44 & * & * & * & $\Omega_{ccc}^{++}$%
\end{tabular}
\end{table}

\narrowtext
\begin{table}[tbp]
\caption{Magnetic moments of spin 1/2 baryons, compared with experimentally
found values. The quantities are theoretically determined by the constituent
quark mass ratios, $m_n/m_s\simeq 0.79,m_n/m_c\simeq 0.23.$ We have included
a few charmed states although the magnetic moment data for them are not yet
available---denoted by ?. When no errors are quoted they are very small.}
\begin{tabular}{c|c|c}
Baryon & Theory & Experiment \\
\tableline p & 2.75 & 2.79 \\
n & -1.84 & -1.91 \\
$\Lambda$ & -0.72 & $-0.61\pm .01$ \\
$\Sigma^+$ & 2.69 & $2.46\pm .01$ \\
$\Sigma^0-\Lambda$ & -1.59 & $-1.61\pm .08$ \\
$\Sigma^-$ & -0.98 & $-1.16\pm .02$ \\
$\Xi^0$ & -1.58 & $-1.25\pm .01$ \\
$\Xi^-$ & -0.66 & -0.65 \\
$\Lambda_c^+$ & 0.20 & ? \\
$\Sigma_c^{++}$ & 2.38 & ? \\
$\Sigma_c^+-\Lambda_c^+$ & -1.59 & ? \\
$\Xi_c^+$ & 0.20 & ?
\end{tabular}
\label{Table 4}
\end{table}

\end{document}